\documentclass[aip,prb,twocolumn,showpacs]{revtex4}
\usepackage{graphicx,color}
\usepackage{amsmath}
\usepackage{longtable}
\begin{document}

\title{Effective Coulomb Logarithm for One Component Plasma}

\author{Sergey A. Khrapak\footnote{Also at Joint Institute for High Temperatures, Moscow, Russia; Electronic mail: skhrapak@mpe.mpg.de} }
\date{\today}
\affiliation{Max-Planck-Institut f\"ur extraterrestrische Physik, D-85741 Garching,
Germany}

\begin{abstract}
An expression for the effective Coulomb logarithm in one-component-plasma is proposed, which allows to extend the applicability of the classical formula for the self-diffusion coefficient to the strongly coupled regime. The proposed analytical approximation demonstrates reasonable agreement with previous numerical simulation results. Relevance to weakly screened Yukawa systems (and, in particular, complex plasmas) is discussed.
\end{abstract}

\maketitle

The one-component-plasma (OCP) model is an idealized system of point charges (e.g., ions) immersed in a neutralizing uniform background of opposite charges (e.g., electrons). This model is of considerable interest from the fundamental point of view and is relevant to a wide class of physical systems, including for example laboratory and space plasmas, planetary interiors, white dwarfs, liquid metals, and electrolytes. There are also relations to various soft matter systems such as charged colloidal suspensions and complex (dusty) plasmas. It is conventional to characterize the OCP system with the ion density $n$ and the temperature $T$ by the dimensionless coupling parameter $\Gamma=e^2/aT$, where $e$ is the charge and $a=(4\pi n/3)^{-1/3}$ is the Wigner-Seitz radius. In the limit of weak coupling, $\Gamma\ll 1$, ions form a disordered gas-like state. Correlations increase with coupling and, at $\Gamma\gtrsim 1$, OCP exhibits properties characteristic of a fluid-like phase. The fluid-solid phase transition occurs at $\Gamma_{\rm M}\simeq 172$.~\cite{Hamaguchi}

The familiar expression for the relaxation frequency in ion-ion collisions is~\cite{NRL}
\begin{equation}\label{nu_0}
\nu_0=(3\pi)^{-1/2}\omega_{\rm p}\Gamma^{3/2}\Lambda,
\end{equation}
where $\omega_{\rm p}=\sqrt{4\pi e^2n/m}$ is the plasma frequency, $m$ is the ion mass, and $\Lambda$ is the so-called Coulomb logarithm. The Coulomb logarithm is well defined in the weakly coupled regime ($\Gamma\ll 1$). Within the framework of kinetic theory, the dominant logarithmic term in $\Lambda$ has the form
\begin{equation}\label{CL_eff}
\Lambda\simeq \frac12\ln\left[1+(\lambda_{\rm D}k_{\rm max})^2\right],
\end{equation}
where $k_{\rm max}$ is the upper limit of integration over $k$ (corresponding to the exclusion of a close proximity of the ion, where kinetic theory is not applicable) and $\lambda_{\rm D}=\sqrt{T/4\pi e^2 n}$ is the Debye radius. The proper choice of $k_{\rm max}$ in this regime is the inverse Coulomb radius ($R_{\rm C}=e^2/T$). The ion-ion repulsion is very strong for $r\lesssim R_{\rm C}$ so that no other ions are essentially present in this region and the kinetic approach is inappropriate. This results in the well known expression
\begin{equation}\label{CL_WC}
\Lambda\simeq\ln\left(\lambda_{\rm D}/R_{\rm C}\right) = -\ln(\sqrt{3}\Gamma^{3/2}).
\end{equation}
The contribution to the relaxation rate from the close proximity of the ion can be estimated as $\nu_1\sim n\sigma v_{T}\sim n (e^2/T)^2\sqrt{T/m}\sim \omega_{\rm p} \Gamma^{3/2}$, where $\sigma\simeq \pi R_{\rm C}^2$ is the relevant scale of the momentum transfer cross section and $v_T=\sqrt{T/m}$ is the thermal velocity. Since $\nu_0/\nu_1\sim \Lambda$, the kinetic approach delivers sufficient accuracy when $\Lambda\gg 1$ or, equivalently, $\Gamma\ll 1$.

Elementary theory of diffusion in weakly coupled regime (binary collisions dominate) relates the self-diffusion coefficient $D$ to the relaxation frequency $\nu$ via $D=T/(m\nu)$. Using Eq.~(\ref{nu_0}) and conventional normalization, $D_*=D/\omega_{\rm p}a^2$, we obtain
\begin{equation}\label{D1}
D_*=\sqrt{\frac{\pi}{3}}\frac{1}{\Gamma^{5/2}\Lambda}.
\end{equation}
The equations (\ref{CL_WC}) and (\ref{D1}), which are often referred to as the Chapman-Spitzer (CS) result, are essentially exact in the weakly coupled OCP. More precisely, they provide   ``logarithmic accuracy'' in the sense that other terms ${\mathcal O}(1)$ have been neglected in comparison with the dominant (large) logarithmic terms in $\Lambda$, as just discussed above.
The question should naturally arise, whether the effective Coulomb logarithm (\ref{CL_eff}) can be meaningfully modified in such a way that the applicability of Eq.~(\ref{D1}) is extended into the regime of strong coupling.

%Recently the application of this simple formula has been extended to moderate couplings by essentially changing $\Lambda$ to $1+\Lambda$ under the Coulomb logarithm.~\cite{Daligault}

In this Brief Communication, it is proposed to use the approach of Nordholm,~\cite{Nordholm} who showed a simple way to greatly reduce inaccuracy of the traditional Debye-H\"{u}ckel theory with respect to evaluating thermodynamics properties of moderately and strongly coupled OCP. He essentially recognized that the exponential ion density must be truncated close to a test ion so as not to become negative upon linearization. This resulted in a modification that he called Debye-H\"{u}ckel plus hole (DHH) theory. In the context of the problem addressed here, such a hole (void) around a test ion which contains no other ions, corresponds to the region where kinetic approach should not be applied. Thus, regardless of other possible limitations,~\cite{Lin} a reasonable choice of $k_{\rm max}$ would be simply the inverse hole radius $h^{-1}$. In the weakly coupled regime the result remains unchanged, since $h\simeq R_{\rm C}$ in this case. In the opposite, strongly coupled regime, we have $h\rightarrow a$, the upper possible limit of the hole radius. Using Eq.~(\ref{CL_eff}) we immediately obtain
\begin{equation}
\Lambda \simeq \frac12\ln\left(1+1/3\Gamma\right).
\end{equation}
In the limit $\Gamma\gg 1$ this yields $\Lambda\simeq 1/(6\Gamma)$ and $\nu_0\simeq (1/6\sqrt{3\pi})\omega_{\rm p}\sqrt{\Gamma}$. The contribution to the momentum transfer from the excluded region (hole) can be roughly estimated considering ion scattering from a hard sphere of radius $h\simeq a$. We get $\nu_1\sim n\sigma v_T\sim n a^2 \sqrt{T/m}\sim \omega_{\rm p}/\sqrt{\Gamma}$. Thus, in the strongly coupled regime, we have $\nu_0/\nu_1\sim \Gamma\gg 1$, i.e. kinetic contribution from distant collisions should again provide dominant contribution to the relaxation rate.

\begin{figure}
\includegraphics[width=7.5cm]{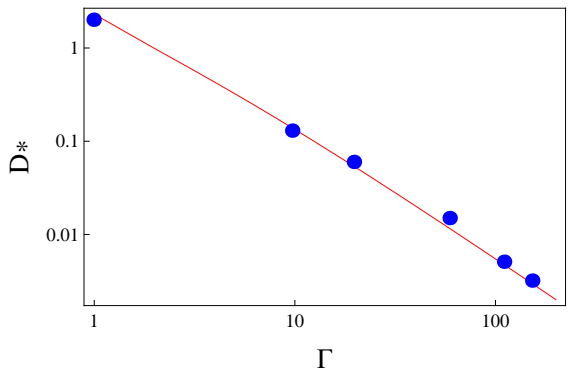}
\caption{(color online). Reduced self-diffusion coefficient $D_*$ of the one-component-plasma versus the coupling parameter $\Gamma$. Circles: Molecular dynamics results.~\cite{Hansen} Curve: Analytical model of Eqs. (\ref{D1}), (\ref{h}), and (\ref{CL}).}
\label{comparison1}
\end{figure}

If this (to some extent naive) picture is relevant, there is a hope that a proper interpolation between the weakly and strongly coupling limits would adequately reproduce the effective momentum transfer rate (\ref{nu_0}) and self-diffusion coefficient (\ref{D1}) in a wide range of $\Gamma$. We proceed by using the original result by Nordholm~\cite{Nordholm} for the hole radius
\begin{equation}\label{h}
h/\lambda_{\rm D}=\left[1+(3\Gamma)^{3/2}\right]^{1/3}-1,
\end{equation}
and postulating the following expression for the {\it effective} Coulomb logarithm
\begin{equation}\label{CL}
\Lambda_{\rm eff}=\frac12 \ln\left[ 1+ (\lambda_{\rm D}/h)^2\right].
\end{equation}
Comparison between the present analytical result of Eqs.~(\ref{D1}), (\ref{h}), (\ref{CL}) and previous molecular dynamics simulation is shown in Figure~\ref{comparison1}. The agreement is quite reasonable, especially taking into account the approximate character of the present model. Although, more accurate fits of the self-diffusion coefficient in Yukawa (including limiting case of the OCP) systems have recently become available,~\cite{DaligaultPRE} they treat separately weakly coupled and strongly coupled regimes, while the present model provides reasonable fit over the entire region of coupling, up to the fluid-solid transition.

\begin{figure}[b]
\includegraphics[width=7.5cm]{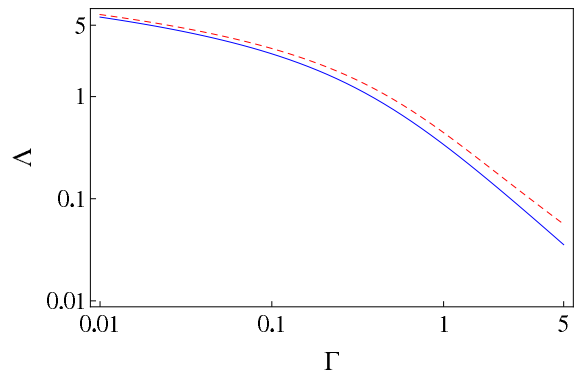}
\caption{(color online). Coulomb logarithm $\Lambda$ versus the coupling parameter $\Gamma$. The solid (blue) curve is a numerical fit to the MD simulation data from Ref.~\onlinecite{Dimonte}. The dashed (red) curve corresponds to the effective Coulomb logarithm proposed in the present work [Eqs.~(\ref{h})-(\ref{CL})].}
\label{comparison2}
\end{figure}

Moreover, one may expect that the proposed form of the effective Coulomb logarithm (possibly with some modifications) can be useful in a more general context of strongly coupled plasmas. As a demonstration, we refer to the results from molecular dynamics simulation of electron-ion temperature relaxation in a classical Coulomb plasma in the regime $0.03\lesssim \Gamma\lesssim 3$, described in Ref.~\onlinecite{Dimonte}. In this numerical study, positively charged electrons and ions on a neutralizing (negative) background were used to avoid recombination, reassembling classical OCP to some extent. A simple numerical fit to the simulations results in the entire regime investigated  has a form~\cite{Dimonte}
\begin{equation}\label{Dimonte}
\Lambda\simeq \ln\left(1+0.4\Gamma^{-3/2}\right),
\end{equation}
A comparison between $\Lambda_{\rm eff}$ from Eq. (\ref{CL}) and $\Lambda$ from Eq.~(\ref{Dimonte}) is shown in Fig.~\ref{comparison2}. Present approximation (\ref{CL}) somewhat overestimates the fit (\ref{Dimonte}), but the relative deviation (which increases with $\Gamma$) does not exceed $\sim 30\%$ for the highest $\Gamma$ investigated.

As a final remark, let us briefly discuss the relevance of the obtained results in the context of complex plasmas. Complex (dusty) plasmas consist of highly charged micron-size particles in a neutralizing plasma background. Significant interest to these systems is in large part related to exciting possibilities to use them (complimentary to other soft matter systems) to study a broad range of important fundamental processes (e.g., phase transitions, self-organizations, transport, rheology, linear and non-linear waves, etc) as well as with numerous applications (industrial plasma processes, fusion, etc.).~\cite{Chaudhuri,Book,Bonitz,FortovPR} The particles in complex plasmas interact with each other via the {\it screened} Coulomb (Debye-H\"{u}ckel or Yukawa) potential (as long as interparticle distance does not exceed several screening lengths).~\cite{KhrapakCPP} In addition, the particles are coupled to the surrounding medium (mainly to the neutral gas, since the plasma ionization fraction is normally quite low) and this coupling can be varied in a broad range. At first glance, there can be very little in common between the particle diffusion in complex plasmas and that in the OCP.

\begin{figure}[t!]
\includegraphics[width=7.5cm]{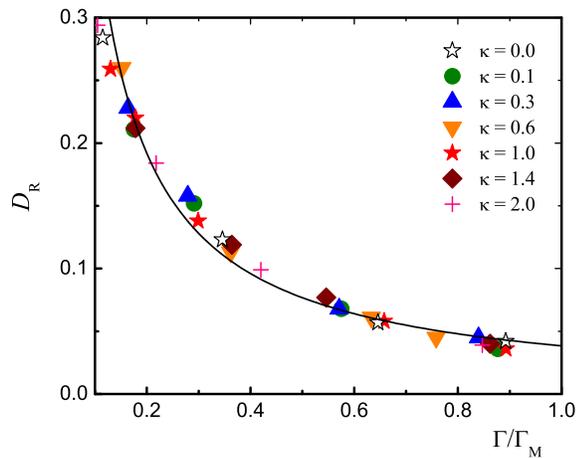}
\caption{(color online) Reduced self-diffusion coefficient of one-component Yukawa systems as a function of the relative coupling strength $\Gamma/\Gamma_{\rm M}$. Symbols: Molecular dynamics results.~\cite{Hansen,Ohta} The solid curve corresponds to the approximation (\ref{D2}) for the strongly coupled regime.}
\label{Yukawa}
\end{figure}

Nevertheless, important similarities do exist. To proceed further it is convenient to introduce another normalization for the self-diffusion coefficient $D_{\rm R}=Dn^{1/3}\sqrt{m/T}$, which is often referred to as Rosenfeld's normalization.~\cite{Rosenfeld} It was shown recently, that in weakly screened Yukawa systems $D_{\rm R}$ is a universal function of the relative coupling strength $\Gamma/\Gamma_{\rm M}$ in the strongly coupling regime.~\cite{KVM} Therefore, the proposed approximation for the OCP, would be sufficient to describe the diffusion in such systems. Using the asymptote $\Lambda\simeq 1/(6\Gamma)$, we easily get $D_{\rm R}\simeq 6.60/\Gamma$. With $\Gamma_{\rm M}\simeq 172$ we obtain
\begin{equation}\label{D2}
D_{\rm R}\simeq 0.0384\left(\Gamma_{\rm M}/\Gamma\right).
\end{equation}
In this form, Eq.~(\ref{D2}) is applicable not only to the OCP, but also to weakly screened Yukawa systems. This is illustrated in Fig.~\ref{Yukawa}. The agreement between the simple expression (\ref{D2}) and numerical data is quite good in the regime $\Gamma \gtrsim 20$ and $\kappa \lesssim 2$, where  $\kappa=a/\lambda_{\rm D}$ is the screening parameter.~\cite{Kappa} Since quite accurate fits for the dependence $\Gamma_{\rm M}$ on $\kappa$ are available,~\cite{Hamaguchi,VaulinaJETP,VaulinaPRE,KhrapakPRL} Eq.~(\ref{D2}) represents a useful tool for estimating self-diffusion in weakly screened Yukawa systems (e.g. complex plasmas in the regime of weak damping).

The effect of damping due to the frictional coupling between the particles and the neutral gas should also be discussed. A simple heuristic expression relating the undamped (Newtonian) and overdamped (Brownian) diffusivities has been recently proposed.~\cite{Pond} In the strongly coupled regime, both reduced diffusivities are proportional to each other, and the relation is particularly simple $D/D_0\simeq c_1 D_{\rm R}$, where $D$ is the actual diffusion coefficient, $D_0=T/(m\nu_{\rm fr})$ is the bare Brownian diffusion coefficient  (no interparticle interactions), $\nu_{\rm fr}$ is the macroscopic friction rate due to collisions between the particles and neutral atoms, and $c_1 = 3.3176$ is a numerical constant.~\cite{Pond} This relation can be used to evaluate the particle diffusion coefficient in complex plasmas in the limit of strong damping (e.g., high neutral gas pressures). Moreover, it has been suggested that in the regime of intermediate damping, the reduced diffusion coefficient can be universally expressed via the dimensionless ''damping index''.~\cite{KVM} This quantity is defined as the ratio between the typical interparticle spacing and mean ballistic free path of the particles $\xi=\nu_{\rm fr}/(n^{1/3}v_T)$ and is a natural measure of the damping strength. The expression proposed in Ref.~\onlinecite{KVM} reads
\begin{equation}
D/D_0\simeq D_{\rm R}\frac{c_1\xi}{c_1+\xi}.
\end{equation}
It is expected to describe particle diffusion in weakly screened strongly coupled complex plasmas in the entire range of frictional dissipation.

To conclude, a heuristic expression for the effective Coulomb logarithm for the one-component-plasma model has been proposed. This expression can be used to describe the self-diffusion in the OCP fluid in the entire regime of coupling strengths below crystallization. It yields essentially exact results in the limit of weak coupling, renders quite accurate results in the limit of strong coupling (near the crystallization point) and provides reasonable approximation in the intermediate regime. In addition, the obtained results can be of certain value in the context of strongly coupled plasmas and weakly screened Yukawa systems (complex plasmas).


\begin{thebibliography}{99}
\bibitem{Lin} These can be for instance associated with the requitements to linearize the kinetic equation. This is not discussed in detail here, since the plausibility of the final result will be checked against ``exact'' results from numerical simulations.
\bibitem{Kappa} In complex plasmas, the interparticle separation is often characterized by $\Delta = n^{-1/3}$ and the screening parameter is defined as $\kappa=\Delta/\lambda_{\rm D}$. In this notation, the condition of weak screening becomes $\kappa\lesssim 3.2$ ($\Delta\simeq 1.6 a$), which is satisfied under many realistic experimental conditions.
\end{thebibliography}
\end{document}